\documentclass{elsart}
\usepackage[english]{babel}
\usepackage{cite}
\usepackage{multirow}
\usepackage{rotating}
\usepackage{color}
\usepackage{graphicx}
\usepackage{epstopdf}

\begin{document}
\begin{frontmatter}

\title{Interevent time distributions  of human multi-level activity in a virtual world}
\author[ICMP]{Olesya Mryglod},
\author[MUW]{Benedikt Fuchs},
\author[MIT]{Michael Szell},
\author[ICMP]{Yurij Holovatch},
\author[MUW,SF,IIASA]{Stefan Thurner\corauthref{cor}}
\corauth[cor]{Corresponding author.}
\ead{stefan.thurner@meduniwien.ac.at}

\address[ICMP]{Institute for Condensed Matter Physics,
               National Acad. Sci. of Ukraine,
               79011 Lviv, Ukraine}
\address[MUW]{Section for Science of Complex Systems, Medical University of Vienna, Vienna, Austria}
\address[SF]{Santa Fe Institute, Santa Fe, NM 87501, USA}
\address[IIASA]{IIASA, Schlossplatz 1, A-2361 Laxenburg, Austria}
\address[MIT]{Senseable City Laboratory, Massachusetts Institute of Technology, Cambridge, Massachusetts, USA}
\begin{abstract}
Studying human behaviour in virtual environments provides extraordinary opportunities
for a quantitative analysis of social phenomena with levels of accuracy that approach those
of the natural sciences. In this paper we use records of player activities  in the massive multiplayer
online game {\em Pardus} over 1,238 consecutive days, and analyze dynamical features of sequences of actions of players.
We build on previous work were temporal structures of human actions of the same type were quantified,
and extend provide an empirical understanding of human actions of different types.
This study of multi-level human activity can be seen as a dynamic counterpart of static multiplex network analysis.
We show that the interevent time distributions of actions in the {\em Pardus} universe follow highly non-trivial distribution
functions, from which we extract action-type specific characteristic ``decay constants''.
We discuss characteristic features of interevent time distributions, including periodic
patterns on different time scales, bursty dynamics, and various functional forms on different time scales.
We comment on gender differences of players in emotional actions, and find that while
male and female act similarly when performing some positive actions, females are slightly faster
for negative actions.
We also observe effects on the age of players: more experienced players are generally faster in making decisions
about engaging and terminating in enmity and friendship, respectively.
\end{abstract}
\end{frontmatter}

\section{Introduction}\label{I}

The life of humans can be viewed as a sequence of different actions that are carried out from birth to death.
Some of these actions are carried out with high regularity on various timescales (circadian, yearly, etc.),
others have significant stochastic components.
By now it is well established that distribution functions characterising sequences of human actions over time are highly non-trivial
\cite{Kleinberg,Barabasi05a,Oliveira05,Vazquez06,Zhou08,Stouffer06,Malmgren09,Goh08,Wu10,Jo12,Yasseri12a,Yasseri12b},
and their origins remain largely unclear.

The very fact that distributions of human actions over time, such as distributions of phone calls, follow statistical laws was known
since the beginning of the last century \cite{Erlang17}, and triggered the origin of queueing theory \cite{queue}. Early attempts to understand
human action sequences were based on the assumption that actions are carried out homogeneously in time with constant rates,
which then lead to Poisson processes \cite{Haight67}. This lead to the conclusion that times between consecutive actions of the same
individual should be distributed exponentially. Models to describe human dynamics that are based on Poisson processes are still widely used  \cite{Anderson91,Malmgren08}. However, the careful analysis of various patterns of human activity provides growing evidence for highly
inhomogeneous bursty distributions of human actions in time.
Further there is evidence for non-trivial inter-dependencies of actions that
influence their (usually power-law dominated) temporal statistics \cite{Kleinberg,Barabasi05a,Oliveira05,Vazquez06,Zhou08,Stouffer06,Malmgren09,Goh08,Wu10,Jo12,Yasseri12a,Yasseri12b}.
The latter were found both for traditional human actions such as writing letters \cite{Oliveira05,Malmgren09},
checking out books in libraries, performing financial transactions \cite{Vazquez06}, or writing e-mails \cite{Kleinberg,Barabasi05a,Stouffer06},
web browsing \cite{Vazquez06}, sending text messages \cite{Wu10}, editing Wikipedia pages \cite{Jo12,Yasseri12b}, and many more.
Different reasons have been suggested to explain their appearance. In particular, the priority queueing
model \cite{Barabasi05a} is a possibility to explain the bursty mechanism of human
behavior by a decision-based queueing process where individuals perform tasks according to some priority.
This may explain the observed correspondence patterns of Darwin and Einstein \cite{Oliveira05}.
Alternatively it has been hypothesized that human correspondence is driven not by responses to
others but by the variation in an individual's communication needs over the course of their lifetime \cite{Malmgren09}.

The common feature of the above mentioned observations is that they are based on the analysis of human actions of a single type:
{\em either} writing e-mails, {\em or} letters, {\em or} making phone calls, etc. The next step beyond these studies is to consider
the much more involved situation where tasks of different types are performed. What is the action dynamics of an individual
(or of a group of individuals) that writes {\em both} e-mails, {\em and} letters, {\em and} makes phone calls? Carrying out actions of
different types we call {\em multi-level human activity}. The analysis of  multilevel activity encounters obvious
difficulties, typically requiring serious efforts on data collection \cite{Lehmann}.
In this study we circumvent this problem by studying the log-files of all actions performed in a virtual world,
where actions of different types are performed by thousands of people \cite{Lazer09,Bainbridge}.
The virtual universe of the online game \emph{Pardus} \cite{Pardus_web} is briefly outlined below.

\begin{figure}[!h]
\vspace{20ex}
\centerline{\includegraphics[width=0.90\textwidth]{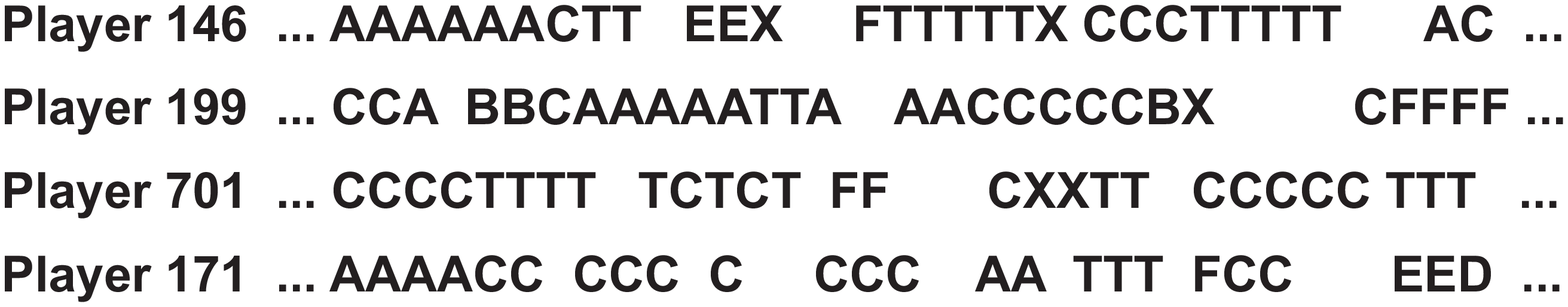}}

\caption{Short segment of action sequences, performed by four {\em Pardus} players.
Different actions are shown by different letters as  explained in Section \ref{II}. The detailed analysis of
time sequences of these multi-level action streams is the main goal of this paper.
\label{fig0}}
\end{figure}

{\em Pardus} is a massive multiplayer online game (MMOG) which is online
since 2004. It is an open-ended game with a worldwide player base of
more than 400,000 registered players \cite{Pardus_web,Szell10a}. The
game has a science fiction setting and features three  different
universes. All universes have a fixed start date but no scheduled
end date. Every player controls one avatar, called the player's
character. The characters act within a virtual world making up their
own goals and interacting with the self-organized social
environment. There is a variety of different activities the
characters can participate in, including communication, trade,
attack, and other forms of social actions such as establishing or
breaking friendships or enmities, see Fig.~\ref{fig0}. Since it has been
launched, the {\em Pardus} game served as a unique testing ground to
measure different observables that characterize inhabitants of the
virtual world and in this way to learn about complex social
processes taking place in the real world -- ``When the same six
soldiers take out a dragon in a synthetic world, the dragon is not
real but the teamwork is'' \cite{Castronova05}. Indeed, with a
complete record of complete information about millions of actions of
different kinds performed by thousands of people during several
years from a single source, this setting provides the unique
position to achieve a detailed non-intrusive quantitative analysis
of complex social behavior
\cite{Szell10a,Szell10b,Szell12,Thurner12,Szell13,Klimek13,Corominas13,Fuchs14a,Fuchs14b,Sinatra14}.
In particular, the complex network structure of the Pardus society
has been exposed and evidence was collected for several
social-dynamics hypothesis, including the Granovetter weak ties
hypothesis and triadic closure \cite{Klimek13}. In this way further
evidence has been provided about validity of a model of online
communities for human societies, allowing to operate with a
precision resembling that of natural sciences \cite{Szell10a}.
Further analysis of the Pardus society has revealed the network
topology of social interactions \cite{Szell10b,Klimek13}, mobility
of characters \cite{Szell12}, behavioral action sequences
\cite{Thurner12}, gender differences in networking \cite{Szell13},
and the functioning of the virtual economy \cite{Fuchs14a}.

The main idea of this paper is to analyze the evolution of actions within the {\em Pardus} universe over time:
what are the temporal characteristics of actions taken by players? Is there an action-specific dynamics, and
are there some global processes that dominate the dynamics of players in the online world? It is well
established by now that a number of features of the virtual society resemble those in the offline society, giving hope to
expect that the analysis of the virtual world action dynamics will provide insights into human dynamics that is
valid also for real societies.
Social systems can be quantified comprehensively by studying the superposition of its constituting socio-economic
networks (multiplex networks, see \cite{multiplex_1,multiplex_2,multiplex_3})
\cite{Szell10b}.  The same information is contained in the dynamical behavioural sequences (Fig.~\ref{fig0}) of the individuals, however
with a focus on temporal aspects of the multi-level  activity.

The paper is organized as follows. In Section \ref{II} we describe the database and concentrate on the main observable of  interest,
the interevent time $\tau$ and its statistics. We comment on the bursty dynamics of individual characters.
Action-specific dynamics is studied in Section \ref{III}. There we analyze inherent features of actions of different types and
compare actions performed by different types of players such as male and female.
The dynamics of the entire  community of players in the {\em Pardus} game is analyzed in Section \ref{IV}. In particular, we
discuss an increase of activity during periods of war in the game. Conclusions and outlook are presented in Section \ref{V}.

\section{Data set and general statistics of interevent times}\label{II}

The {\em Pardus} world consists of three different game universes: Orion, Artemis, and Pegasus. In this study we concentrate on the Artemis
universe because in this Universe complete data is available for all actions of all players \cite{Szell10a}. The Artemis universe is also
the most densely populated universe over time, inhabited by more
than 7,000 active characters (see Table \ref{tab1} for more
details). Artemis was opened on June 10, 2007. Our analysis is based
on the information about player activities during 1,238 consecutive
days since the universe was opened. Every character in the game is a
pilot who owns a spacecraft, travels in the universe and is able to
perform a number of activities of different type. There is no
specified goal in the game, the players make up their own goals and interact with their self-organized
social environment.
Every player picks a male or female avatar, the age of the player is measured as a the number of days
since the first registration.
We concentrate on the following actions that can be acted out within the
game
\begin{itemize}
\item sending private messages from one player to another (communication, C);
\item attacking other players or their belongings (attack, A);
\item trading or giving gifts (trade, T);
\item marking friends by adding their names to a friend list (F);
\item marking enemies by adding their names to an enemy list  (E);
\item removing friends from the friend list (D);
\item removing enemies from the enemy list (X).
\end{itemize}

More information is found in \cite{Szell10a}. It is important to note that actions performed
by a character (such as trade, travel, adding/removing friends) cost a certain amount of so-called
action points (APs). The number of APs available per character at once cannot exceed 6,100.
For each character which owns less APs than their maximum, every 6 minutes 24
APs are automatically generated. Once a character is out of APs,
they have to wait for new APs to be regenerated. Social interactions, e.g. sending private
messages, planning, coordination, etc., do not cost APs. This feature adds one additional
difference between the action types and causes peculiarities in the
behavior of characters, as we will see below.

The above actions can be labeled as ``positive'' or
``negative'' depending on their nature: A, E and D are considered as
negative, while C, T, F, X are considered positive \cite{Szell10a,Thurner12}. In some cases this
classification may be questioned. For example, thorough
categorization of the private messages would have to take into
account their content. However, we will classify C as a positive
action because only a negligible part of communication in Pardus
takes place between enemies \cite{Szell10a}. Since private messages are the
most frequently performed actions ($\approx 80\%$), the number of positive actions
is much larger than the number of negative ones.

\begin{table}[t]
\caption{Characteristics of the data set.
$N$ is the number of characters that performed actions of a given type; $N_a$, overall number of actions performed by these characters
during the observation period within $1,238$ consecutive days. $N_\tau$ is the
number of interevent time ($\tau$) values.  $\tau^{\rm min}$,
$\tau^{\rm max}$ are minimal and maximal values of interevent times given in
seconds.
\label{tab1}}
\vspace{2ex} {\small{ \centerline{
\begin{tabular}{|c|r|r|r|r|c|}
\hline Action type
&$N$& $N_a$& $N_\tau$ & $\tau^{\rm max}$ & $\tau^{\rm min}$ \\
\hline All &7,818&8,373,209&8,365,391&63,882,774&0\\
\hline Positive & 7,806&7,492,460&7,484,648&63,882,774&0\\
\hline Negative &6,918&880,749&873,831&91,715,678&0\\
\hline A &5,412&742,798&737,386&100,686,258&0\\
\hline T &6,883&561,327&554,444&86,705,400&0\\
\hline C&7,799&6,775,950&6,768,151&72,931,246&0\\
\hline E &5,638&105,958&100,320&100,232,580&2\\
\hline F &7,067&125,984&118,917&100,736,362&1\\
\hline X & 3,653&29,199&25,546&87,400,240&1\\
\hline D &3,188&31,993&28,805&95,379,313&1\\
\hline
\end{tabular}}}}
\end{table}

Two observables are widely used to quantify action streams
\cite{Barabasi05a,Oliveira05,Vazquez06,Stouffer06,Malmgren09,Wu10,Yasseri12b,Zhou08_2}.
These are {\em interevent time}, i.e. the time interval between two
consecutive actions of the same person, and the {\em waiting time},
defined as the time interval between the action (of one person) and
reaction (of another person). The last quantity is sometimes
understood as the ``time to reciprocate''.
In the correspondence pattern analysis the reaction on a
received letter is another letter that is sent in reply. However,
there might also be mixed reciprocity, where the type of reaction
differs from an action itself -- say, an email sent as a reaction on
a phone call received. Because of this, we first simply
focus on the interevent time, ignoring the type of
action. We denote this by $\tau$. In Tab. \ref{tab1} we collect the main
features of the data set. The overall number of
characters that have performed at least one action in the Artemis
universe is $N=7,818$, the overall number of actions
performed during the observation period of $1,238$ consecutive days is
$N_a=8,373,209$. The majority of actions belong to the positive type
(F,C,T,X), due to the fact that C are considered as positive
actions. The next in frequency are: A, T, F, E, D, and X.
The number of actions $N_a$ is action type specific and ranges from
$N_a=6,775,950$ for C, to $N_a=29,199$ for X. The number of players
$N$ that are involved in any specific type of action is distributed
rather homogeneously: several thousands of players for each action.

\begin{figure}[!ht]
\centerline{\includegraphics[width=0.7\textwidth]{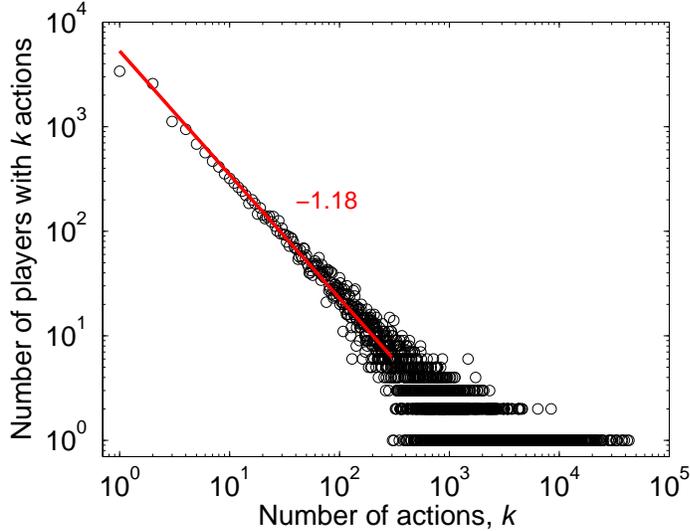} }
\caption{Number of players who performed $k$ actions during the
observation time. The majority of players performed  few
actions during the entire period, while a few very active players
provided huge activity statistics. The line shows scaling with
an exponent of $\sim -1.18$.
\label{fig1}}
\end{figure}

\begin{figure}[t]
\centerline{\includegraphics[width=0.48\textwidth]{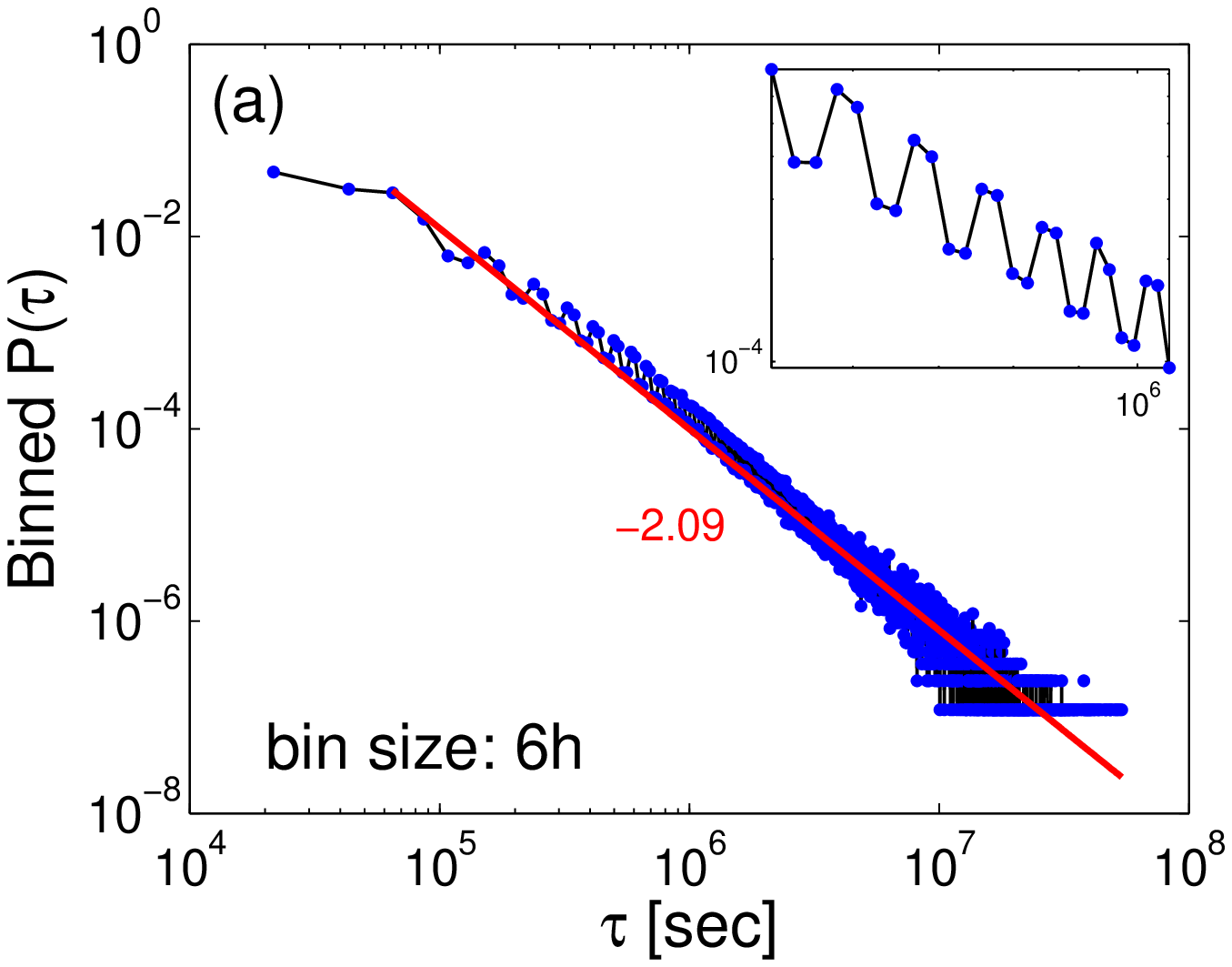}
\includegraphics[width=0.48\textwidth]{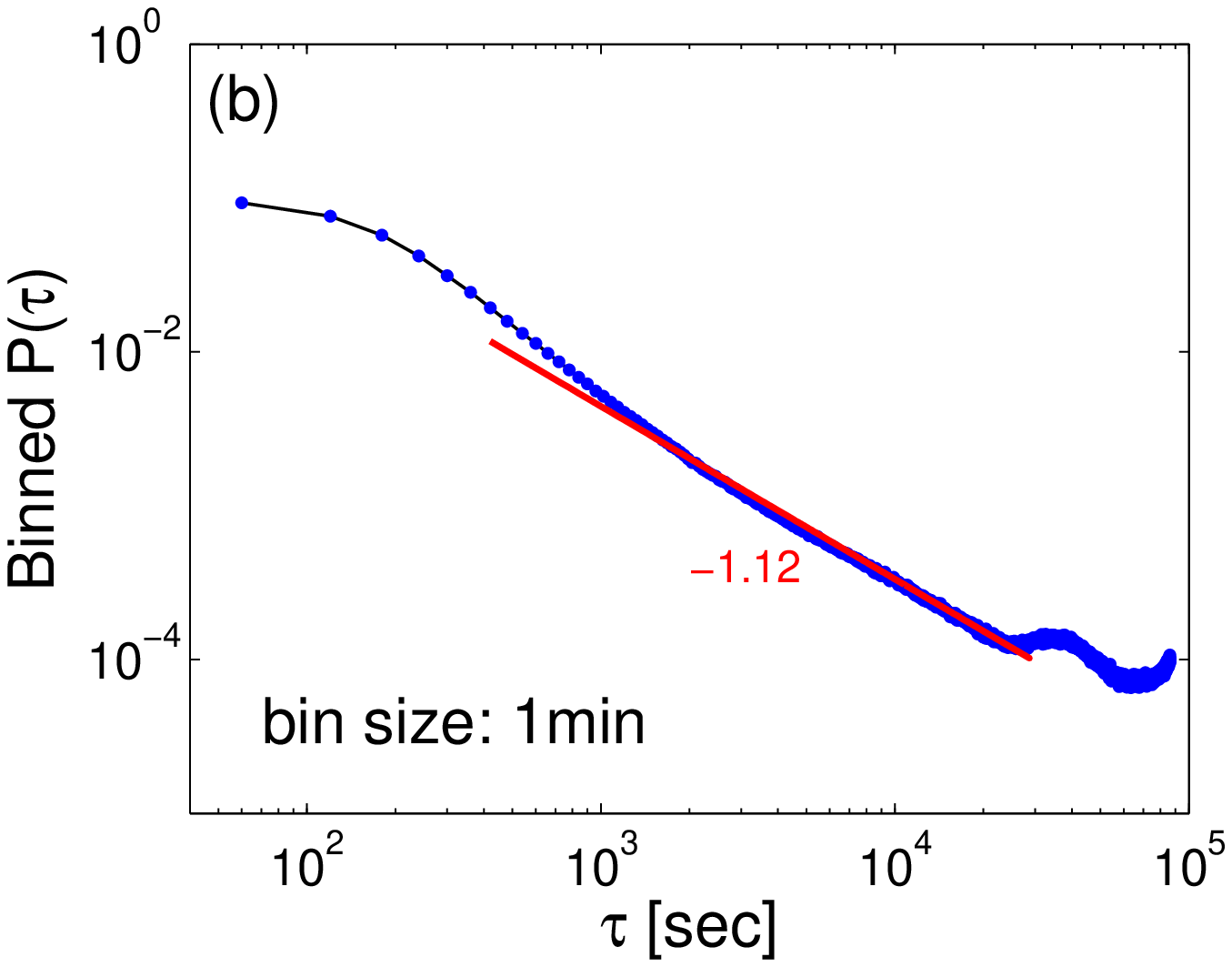} }
\caption{Distribution of the interevent time $\tau$ for all players who performed at least 50 actions.
(a) entire observation period (1,238 days), bin size is 6 hours=21; 600
sec. The inset shows a six day period. Circadian rhythms are clearly visible.
(b) First 24 hours, bin size is 1 min.
\label{fig2} }
\end{figure}

The histogram of the number of actions is shown in Fig.~\ref{fig1}. The majority of players performed just a
few actions during the entire period, while a few very active players provided huge activity statistics. We find an
approximate power law scaling with an exponent of $\sim -1.18$. Players with a small number
of actions typically quit the game just after registration. These players are not representative so that
we exclude all players that performed less than 50 actions.
Since all actions are time stamped with an accuracy of one second,
we can calculate interevent times, $\tau$.
The distribution of $\tau$ follows a power law as seen in
Figs.~\ref{fig2}a and \ref{fig2}b. The exponent dependents on the chosen time scale (bin
size).
The periodic pattern (inset) corresponds to circadian cycles \cite{Jo12,Yasseri12a,Malmgren08}.
%
To focus on the dynamics within the first 24 hours (most frequent values for
$\tau$) we take a bin size of 1  min, Fig.~\ref{fig2}b.
The local minimum, which appears on this scale for $\tau= 7$ hours $=25,200$ sec
can be explained by the ``active working day''. Since it is more convenient to play in the morning (before
work) or in the evening (after work),  interevent intervals of  7 hours are less probable than those for  8, 9, or 10
hours.
One observes a further regime at small values of $\tau$, corresponding to
immediate or slightly delayed repetition. This can be observed within the first 3
minutes after the performed action. This is due to the peculiarities of different
actions. For example, attacks could be naturally grouped into
sequences due to repeatedly pressing an attack button to attack the
same opponent, while a separate decision is needed to add each new
enemy. Further, player synchronization might play a
substantial role, where two players can react promptly when they are
both online, but only with a (significant) delay if one is offline. It is
possible that the shape of the interevent times
distribution after these first ``immediate'' values is influenced by the login behavior of players.

The power-law-like distribution of interevent times between the
actions indicates the bursty nature of human dynamics
\cite{Barabasi05a,Oliveira05,Zhou08}: periods of high activity are
separated by long periods of inactivity. Although the origins of
such a non-uniform distribution of actions are highly diverse, it is
recognized to be an inherent feature of human dynamics. A
measure for burstiness  $B$  was introduced in \cite{Goh08},
its simplified version (see \cite{Jo12,Yasseri12b}) is defined as follows:
\begin{equation}\label{1}
	B \equiv \frac{\sigma-m}{\sigma+m} \quad ,
	\label{eq_burst}
\end{equation}
where $m$ is the average of the interevent time $\tau$, and $\sigma$ its  standard deviation.
For a regular pattern we have $B \sim -1$, for a random Poissonian process with a
fixed event rate, $B \sim 0$, and for fat-tailed distributions of time intervals, $B \sim 1$.

\begin{figure}[t]
\centerline{\includegraphics[width=0.98\textwidth]{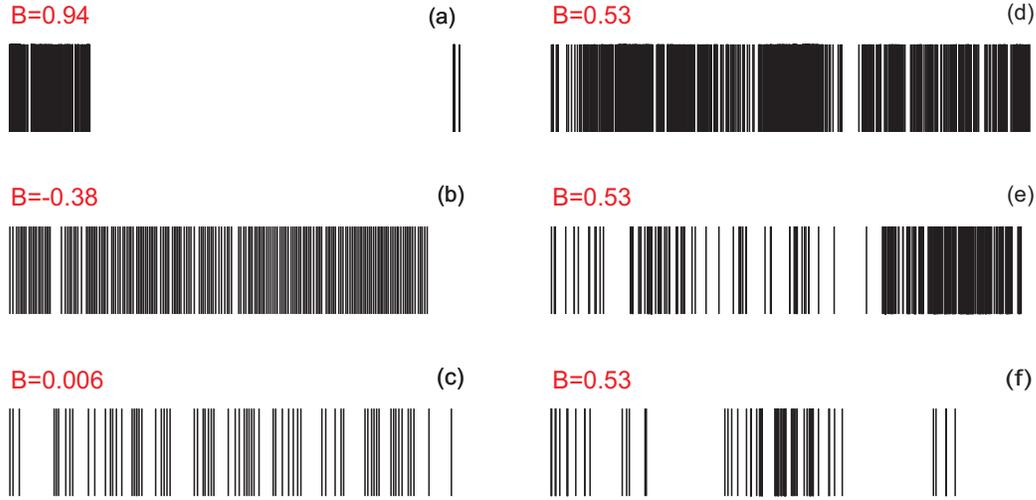}}
\caption{Action streams of players with different values of
burstiness $B$. Lines mark times of  executed actions, the distance between
lines is the interevent time. \label{fig3}}
\end{figure}

\begin{figure}[t]
\centerline{\includegraphics[width=0.8\textwidth]{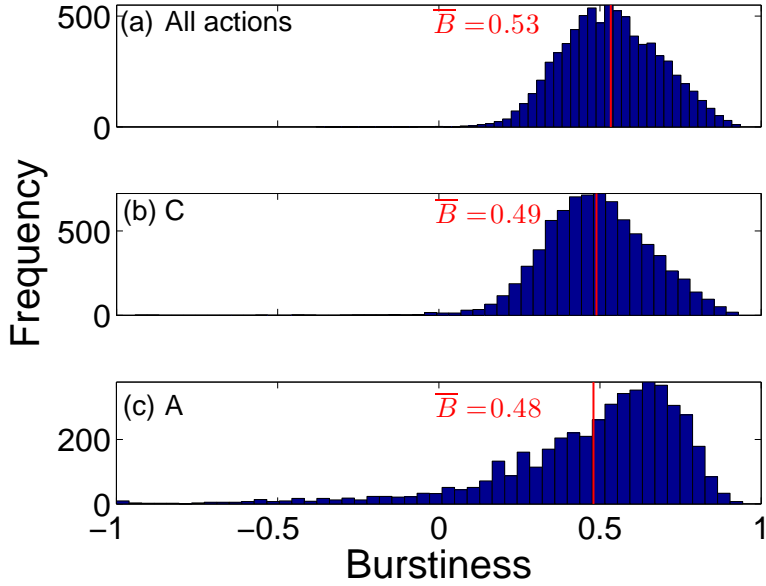}}
\caption{Histogram of burstiness calculated for (a) all action types, (b) communication C, and (c) attacks A.
The most frequent (maximum) ($\hat{B}$) and the mean ($\overline{B}$, red lines) values are {$\hat{B}\simeq
0.53$}, {$\overline{B}\simeq 0.53$} (a); $\hat{B}\simeq 0.50$, $\overline{B}\simeq 0.49$ (b);
$\hat{B}\simeq 0.65$, $\overline{B}\simeq 0.48$ (c), respectively.
\label{fig5}
}
\end{figure}

In Fig.~\ref{fig3} we show (compressed) action streams of players with different values of $B$.
Each line marks the time of an executed action. The patterns on the left-hand side correspond to
the activity with (a) maximal, (b) minimal, and (c) close to zero values of $B$.
A random distribution of actions over the time line is characterized by
$B\sim 0$. It is observed that seemingly very different activity
patterns can be characterized by similar values of burstiness, Fig.~\ref{fig3}d--f.
The histogram of $B$ for all players is shown in Fig.~\ref{fig5}a. The most frequent value (maximum) ($\hat{B}\simeq 0.53$) as well as
the average (mean) ($\overline{B}\simeq 0.53$) values are both larger than 0.5. The average burstiness values for the process of
real-world mobile communication is more than two times smaller, $B\simeq0.2$ \cite{Jo12}, while for Wikipedia editing (events
correspond to  consequent edits of Wikipedia articles) it is $\sim0.6$ \cite{Yasseri12b}.
The maximum of the histogram of $B$,  for attacks, Fig.~\ref{fig5}b, has larger values than for communication, Fig.~\ref{fig5}c.
This illustrates the intuitive understanding of the nature of these actions: attacks appear highly clustered within short time
intervals, while communication is more uniformly distributed over time.

\section{Action-specific dynamics: Decay constants of human actions}\label{III}

In this section, we ask if it is possible to discriminate between actions types given only information about their interevent time distributions.
We can show that the ``decay'' of the interevent time distribution serves as a distinguishing feature of  action types.
To quantify this decay we introduce ``decay constants'' and ``decay exponents" that are specific for different actions.
\begin{figure}[t]
\centerline{\includegraphics[width=0.75\textwidth]{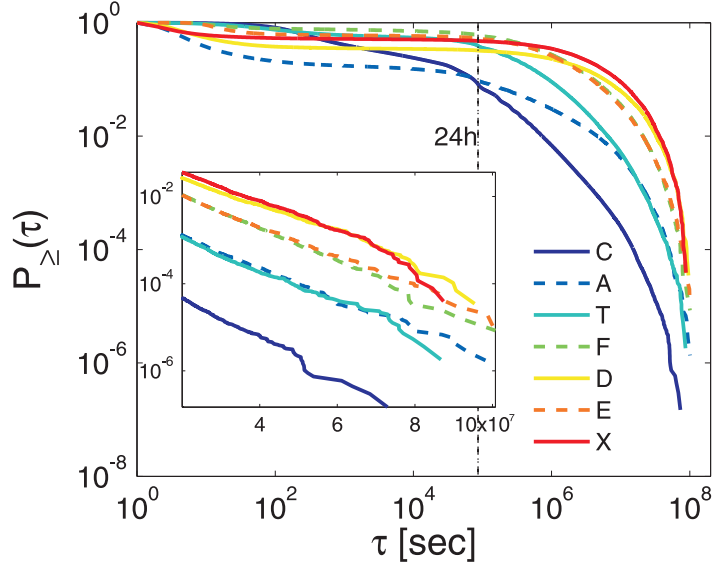}}
\caption{
Inverse cumulative distribution of interevent times $\tau$ for all players
(no binning) and for different kinds of actions.
The actions are: communication (C);
attack(A); giving gifts / trade  (T); making friends/enemies (F/E);
removing friends/enemies (D/X).
Inset: same plot in log-linear scale, for $\tau > 2\cdot10^7$ sec ($>$ 8 months).
\label{fig6} }
\end{figure}
We calculate the inverse cumulative distributions of interevent times $P_{\geq}(\tau)$ for each of the seven action types. They are shown
in Fig. \ref{fig6} for communication (C), attacks (A), trade  (T) and for the friendship-enmity marking actions F, D, E, and X.
Due to the massive contribution of C actions, the curve for {\em all} actions is dominated by the  C distribution.

Based on our previous observations (see Section \ref{II}), we are particularly interested in the behavior on three
different time scales: {\em immediate reaction} where $\tau$ does not exceed a couple of minutes; {\em early day}, $\tau$ is less than 8
hours; and the {\em late day}, $\tau$ is between 8 and 24 hours.
Our main observations are

\begin{figure}[t]
\centerline{
\includegraphics[width=0.45\textwidth]{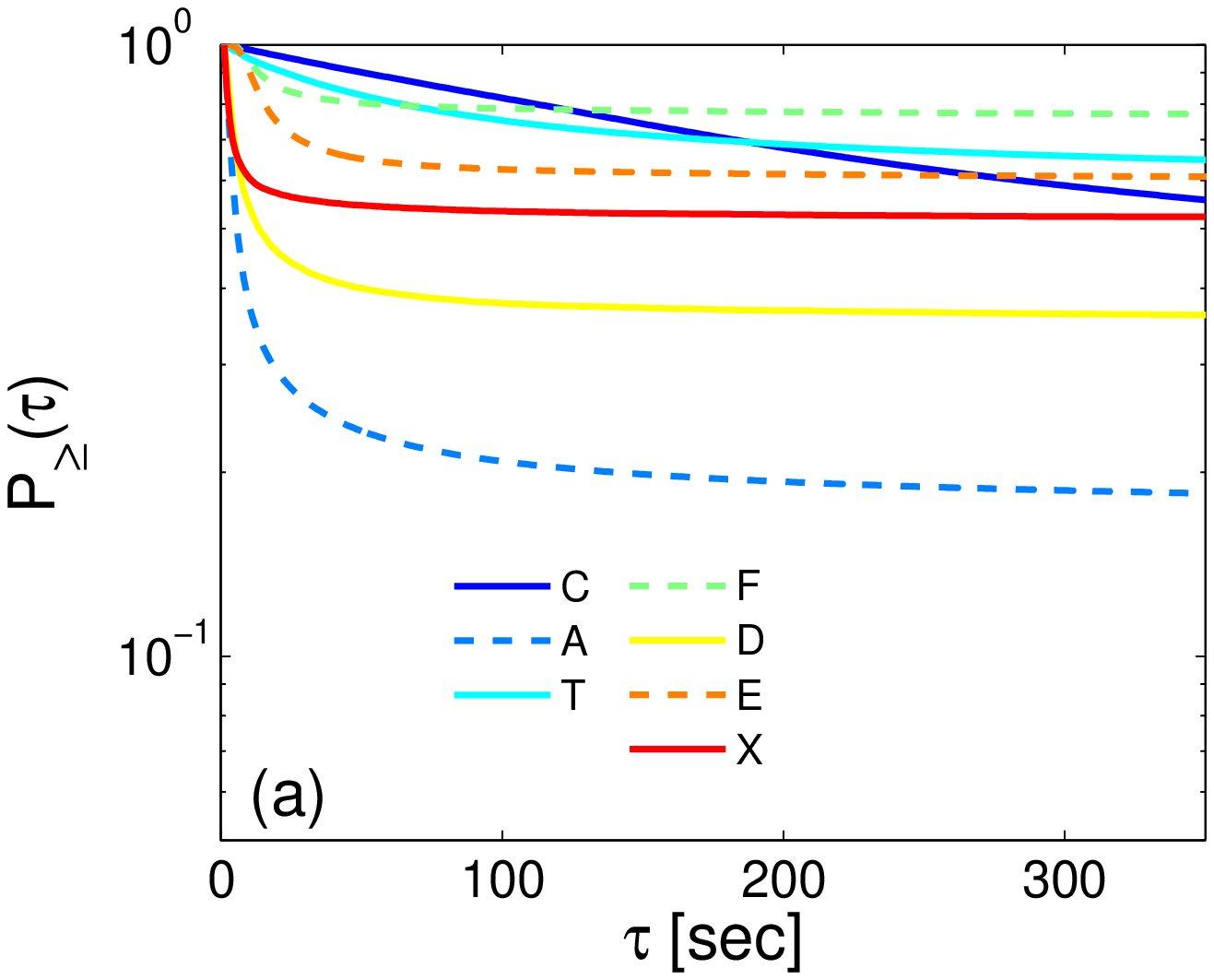}
\includegraphics[width=0.45\textwidth]{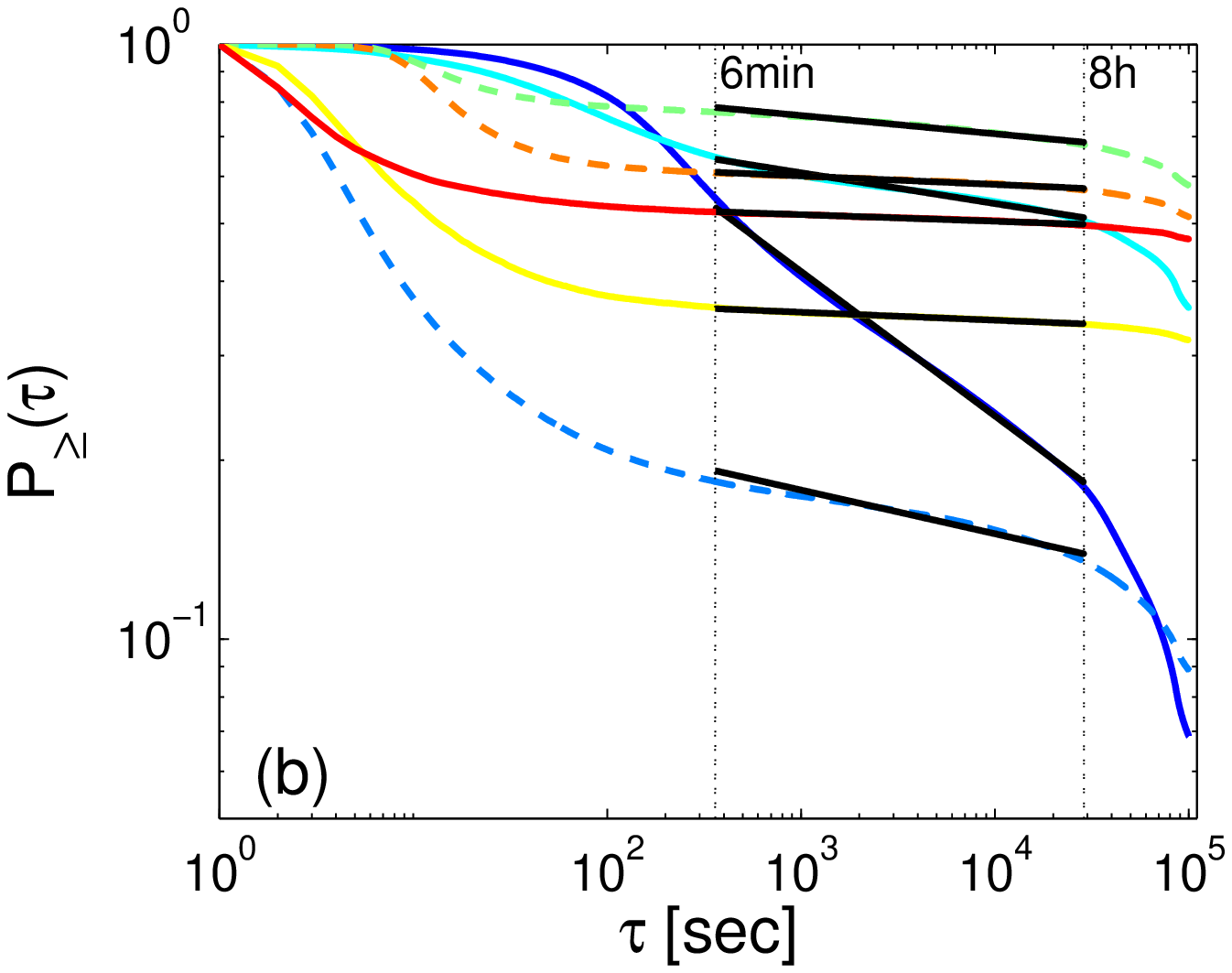}
\includegraphics[width=0.45\textwidth]{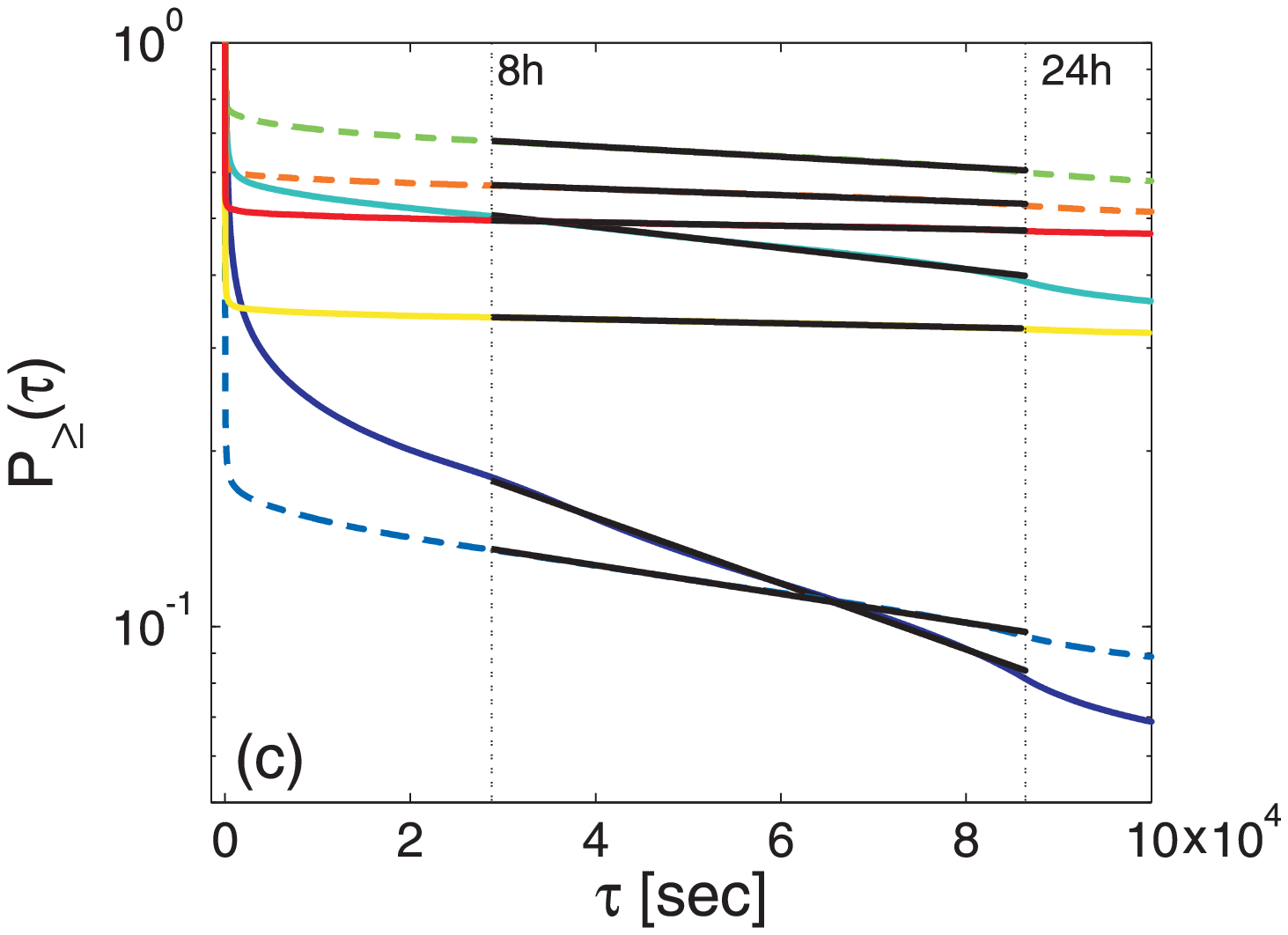}}
 \caption{Inverse cumulative distributions of $\tau$  in different time
 intervals. (a) $\tau\leq 360$ sec, log-linear scale; (b) $\tau\leq 10^5$ sec, log-log scale
 (fits obtained from 6 min -- 8 hours interval); (c) $\tau\leq 10^5$ sec, log-linear scale (fits from 8--24
hours interval).
Behavior on three different time scales is governed by different decay constants, see Tab. \ref{tab1a} and text.}
\label{fig7}
\end{figure}

\begin{itemize}

\item{\em Immediate reaction} ($\tau\leq 360$ sec). $P_{\geq}(\tau)$ is shown  in Fig.
\ref{fig7}a. All curves have a tendency to decay fast at small values of $\tau$. The decay is
especially pronounced for A, D, F, X, and E actions. This means that short interevent times are
typical for most of the actions. Large numbers of attacks or emotional addings-removings are
performed one-by-one in a very fast way (grouped into sequences), while the probabilities of $\tau$
values greater than 1 min start to decrease very slowly. The decay for short interevent
times for T and C is less steep, whereas its further evolution remains more homogeneous.

\item{\em Early day} (6 min $< \tau <$ 8 hours).
For this case $P_{\geq}(\tau)$ is  best approximated by a power law
\begin{equation}\label{2}
	P_{\geq}(\tau) \sim \tau^{-\alpha} \quad ,
\end{equation}
with small exponents in the range of $\alpha \sim 0.01 - 0.1$ for all actions except C, for which we have $\alpha \sim 0.24$,
see Fig.  \ref{fig7}b. The exponents for each curve are collected in the second column of Tab. \ref{tab1a}.

\item{\em Late day} (8 hours $< \tau <$ 24 hours).
$P_{\geq}(\tau)$  in this region is shown in Fig.~\ref{fig7}c. We fit here the decay by the exponential function,
\begin{equation}\label{3}
	P_{\geq}(\tau) \sim \exp{(-\tau/\tau_0)} \quad .
\end{equation}
Values of $\tau_0$  are collected in the third column of Tab.  \ref{tab1a}.
The slow exponential decay is described by slightly different action-specific values of $\tau_0$.
This difference has a tendency to diminish for larger time intervals. The fastest decay is
observed for C, with $\tau_0\simeq 5.6\cdot10^{-6}$, which is twice as large
as for A ($\tau_0\simeq 2.5\cdot10^{-6}$), and ten times larger than for the other actions.

\item{\em Long times}Ê($\tau>2\cdot 10^{7}$ sec).
An exponential cut-off for large $\tau$ (more than eight months) becomes apparent in $P_{\geq}(\tau)$, see
Fig.~\ref{fig6}. The exponential decay has similar numerical values for the different actions providing
evidence for a cut off effect.

\end{itemize}

The results indicate that the various types of actions are characterised by
distinct decay properies of the interevent time distributions.
The values $\alpha$ in Eq. (\ref{2}) and $\tau_0$ in Eq. (\ref{3}) may serve as ``decay constants''  and
turn out to be action specific.
This is especially pronounced for interevent times in the time intervals ``early-day'' (6 min $< \tau <$ 8 hours), and
``late-day'' (8 hours $< \tau <$ 24 hours). The ``long time'' exponential decay can be safely interpreted as a finite
size effect of the sample.
The  average values of interevent times $\overline{\tau}$ also point on the different speed in
performing different kinds of actions. Due to the contribution of some rare but large values of $\tau$, as well
as the periods of natural inactivity (circadian cycles), the values of $\overline{\tau}$ are large:  $\sim 12.6$h for C,
$\sim 51.8$h for A, $\sim 117.1$h for T, $\sim 439.3$h for F,  $\sim 579.3$h for D, $\sim 414.3$h for E, $\sim 829.5$h for X.
$\overline{\tau}$ is smallest for communication and much larger for any other kind of activity.

\begin{figure}[t]
\centerline{\includegraphics[width=0.55\textwidth]{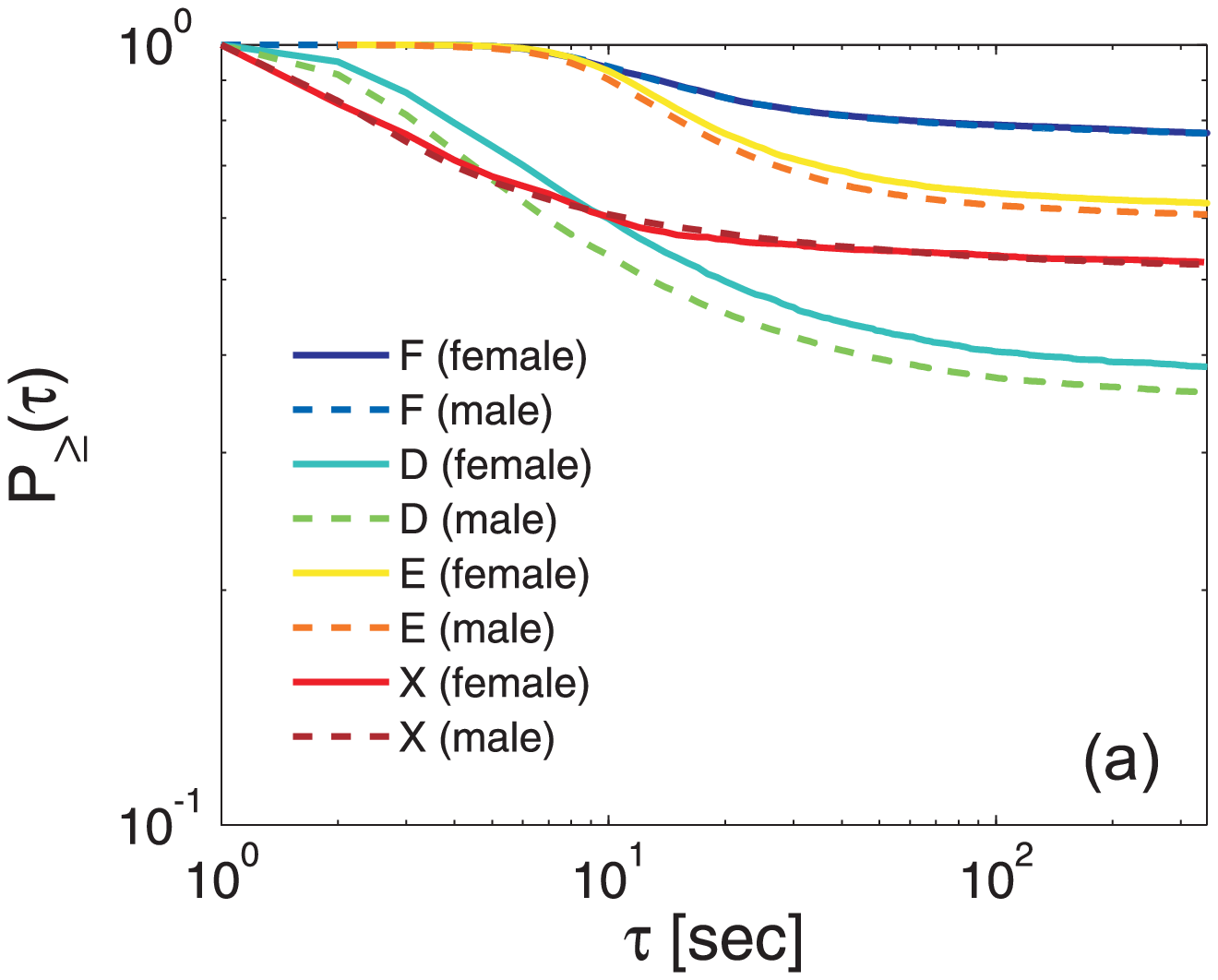}
\hfill
\includegraphics[width=0.55\textwidth]{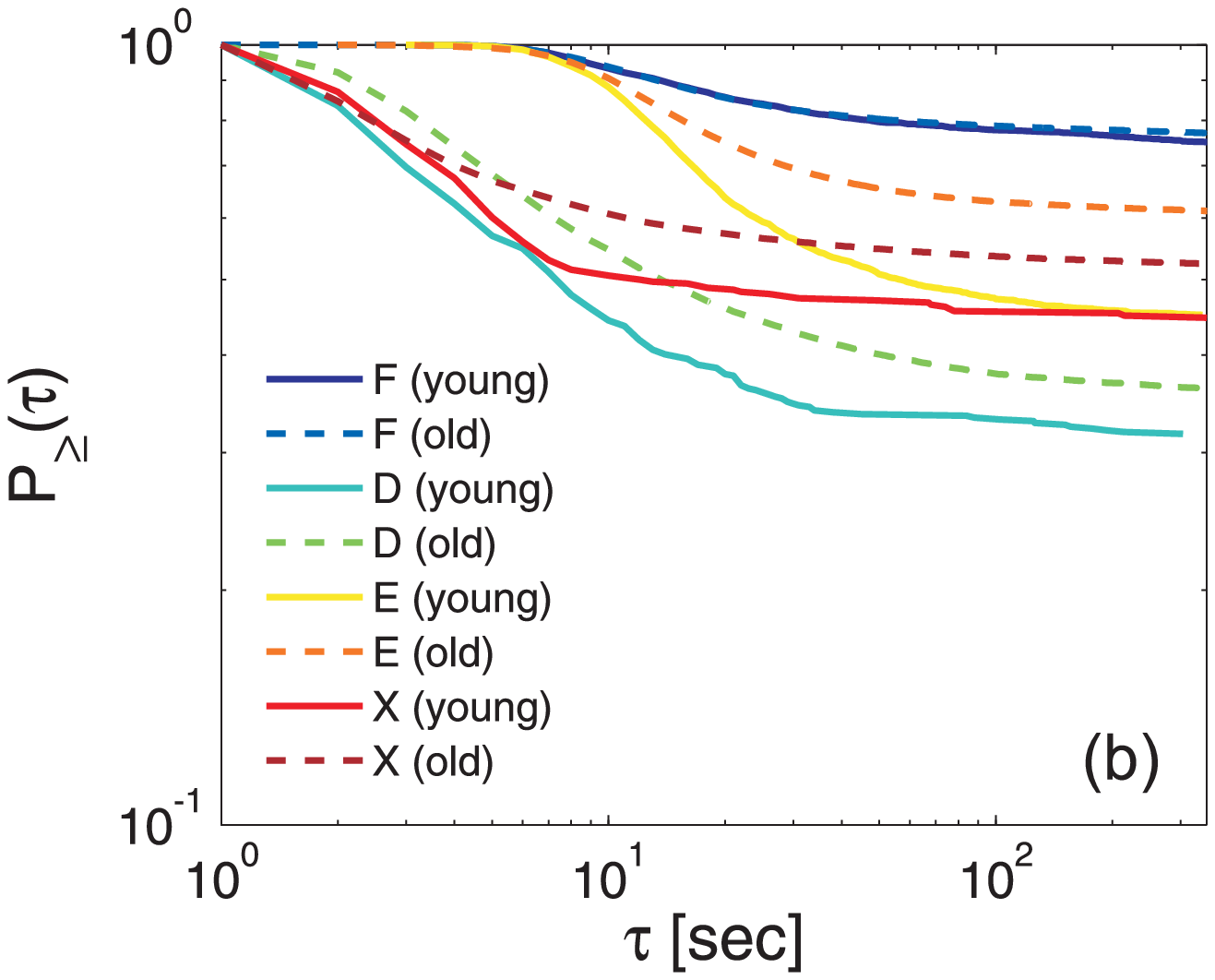}}
\caption{
Inverse cumulative distributions $P_{\geq}(\tau)$  (for $\tau\leq 360$ sec) for
different type of players: (a) male (dashed) and female (solid), (b) young (solid) and old (dashed).
Gender differences for actions C, T and A are minimal, we only present  F, E, D and X actions here.
Whereas the interevent time distributions of males and females performing positive actions (F and X) almost
coincide, this is not the case for the negative actions (E and D). The situation is different for the groups of
young and old players: marking of friends is performed with about the same speed for both ages,
the old players are always faster with the other actions.}
\label{fig8}
\end{figure}

We conclude this section by studying players of different type.
For the majority of actions, the interevent time distributions of males
and females are very similar. We concentrate here on the
actions where slight gender effect is present. In
Fig.~\ref{fig8}a we compare $P_{\geq}(\tau)$ (for small $\tau$) for the E, F, D, and X
actions of female (solid curves) and male (dashed curves)
game characters. There is a slight gender effect in action
interevent times in the process of marking friends and enemies,
compatible what has been previously reported on a social network
level in \cite{Szell13}.
The difference for the negative actions (enemy adding and friend
removing) and the positive actions (friend adding and enemy
removing) might be explained by underlying biological or
socio-dynamic reasons.
Whereas the interevent time distributions of males and females performing positive actions (F and X) almost
coincide, this is not the case for the negative actions (E and D), see Fig. \ref{fig8}a.
The corresponding curves for females are always above those for males: negative actions performed by
females have shorter time intervals. Whereas male and female act
almost similarly when performing positive actions, females are
faster in negative ones. Note, however, that such difference is not
observed for the A, T, and C actions.

\begin{table}[t]
\caption{Decay constants  and  decay exponents for various  actions at different time scales.
The interevent time  distribution $P_{\geq}(\tau)$ for the early-day interval is
governed by the power law, Eq. (\ref{2}), whereas the late-day and long-time intervals are governed by the exponential decay,
Eq. (\ref{3}). Values for $\alpha$ and $\tau_0$ are given in the table for different types of actions. \label{tab1a}}
\centerline{
\begin{tabular}{c c c c c}
\hline 
& early-day $\alpha$ & late-day $\tau_0$& long times $\tau_0$\\
%
\hline
 A   & -0.07         & -2.46e-06     &  -3.64e-08   \\
 T   & -0.05         & -1.81e-06     &  -3.71e-08   \\
C   & -0.24         & -5.65e-06     &  -4.55e-08   \\
E   & -0.01         & -0.56e-07     &  -3.49e-08   \\
F   & -0.03         & -0.88e-07     &  -3.80e-08   \\
X   & -0.01         & -0.29e-07     &  -3.41e-08   \\
D   & -0.01         & -0.34e-07     &  -3.13e-08   \\
\end{tabular}} 
\end{table}

The details of the friendship-enmity marking actions can be observed also for
players with different experience. Fig.~\ref{fig8}b shows $P_{\geq}(\tau)$ (for small $\tau$) are shown for young (solid curves) and
 old (dashed curves) players: the F curves coincide for young {\em Pardus} players.
Old players are defined as those who have started the game at least one year before the last day of the data set.
Young players are those who are younger than one year.
The other curves for old players are always above those of the young players. The time intervals between marking somebody as an
enemy or removing the marks are shorter for experienced players. The speed  of marking friends is the same for all players.
The majority of friends are typically marked early on in  {\em Pardus} ``life''.

\section{Global dynamics and activity patterns: War and peace}\label{IV}

\begin{figure}[h]
\centerline{\includegraphics[width=0.8\textwidth]{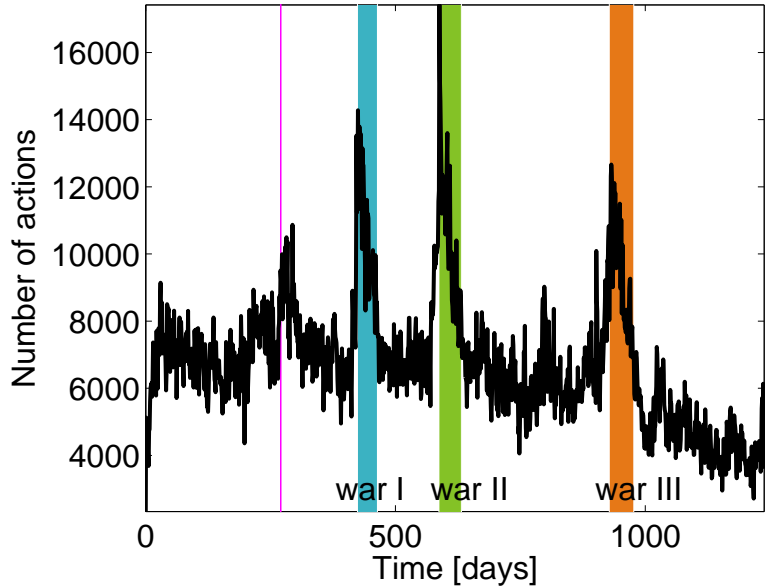}}
\caption{Number of actions (all types) per day on the timeline. There are four pronounced peaks in the player
activities that correspond to specific  events that happened in the virtual world in the observation period:
the three coloured vertical stripes indicate war periods, the vertical line indicates the introduction of a major
new game feature.
\label{fig9}}
\end{figure}

We now study how actions are distributed in time. The number of all actions types pooled together is shown for every day in
Fig. \ref{fig9}. The time interval starts on 2007/06/12 and covers 1,238 consecutive days of players
activity. There are four pronounced peaks in the player activity which can not be understood without knowing the history of
the virtual world during the observation period.
\begin{itemize}
\item The peak of activity on March, 2008 (magenta line in figure)
was caused by the introduction of new major game features called ``Syndicates'' on March 7, 2008 \cite{Syndicates}.
Big changes in the game usually become a hot topic to discuss,  leading to the higher level of C activity.
\item  The peak of activity in August-September, 2008 (blue stripe in figure)
corresponds to the $1^{\rm st}$ war, that occurred in the period between 2008-08-08 and 2008-09-17 (war I).
\item The peak of activity in January-March, 2008 (green stripe in figure)
corresponds to the $2^{\rm nd}$ war, 2009-01-18 -- 2009-03-04 (war II).
\item The peak of activity in end December 2019 -- February 2010
(orange stripe in figure) corresponds to a $3^{\rm rd}$ war between 2009-12-25 and 2010-02-12 (war III).
\end{itemize}

Before  discussing player activity in peace and war periods we briefly explain how a war emerges in the
virtual world. According to the specifics of the game \cite{Pardus_web} each player can belong to one of three {\em factions}
(organizations) or stay neutral. There are three factions in the game: 1 -- ``Federation'', 2 -- ``Empire'', 3 -- ``Union''. Only two
factions can participate in a war simultaneously. During the
observation period the following pairs of factions were involved in wars:
Factions 1 and 2 in war I and factions 1 and 3 in the next two wars.
Factions 2 and 3 were never engaged in a war within the observation period.

\begin{figure}[t]
\centerline{\includegraphics[width=0.55\textwidth]{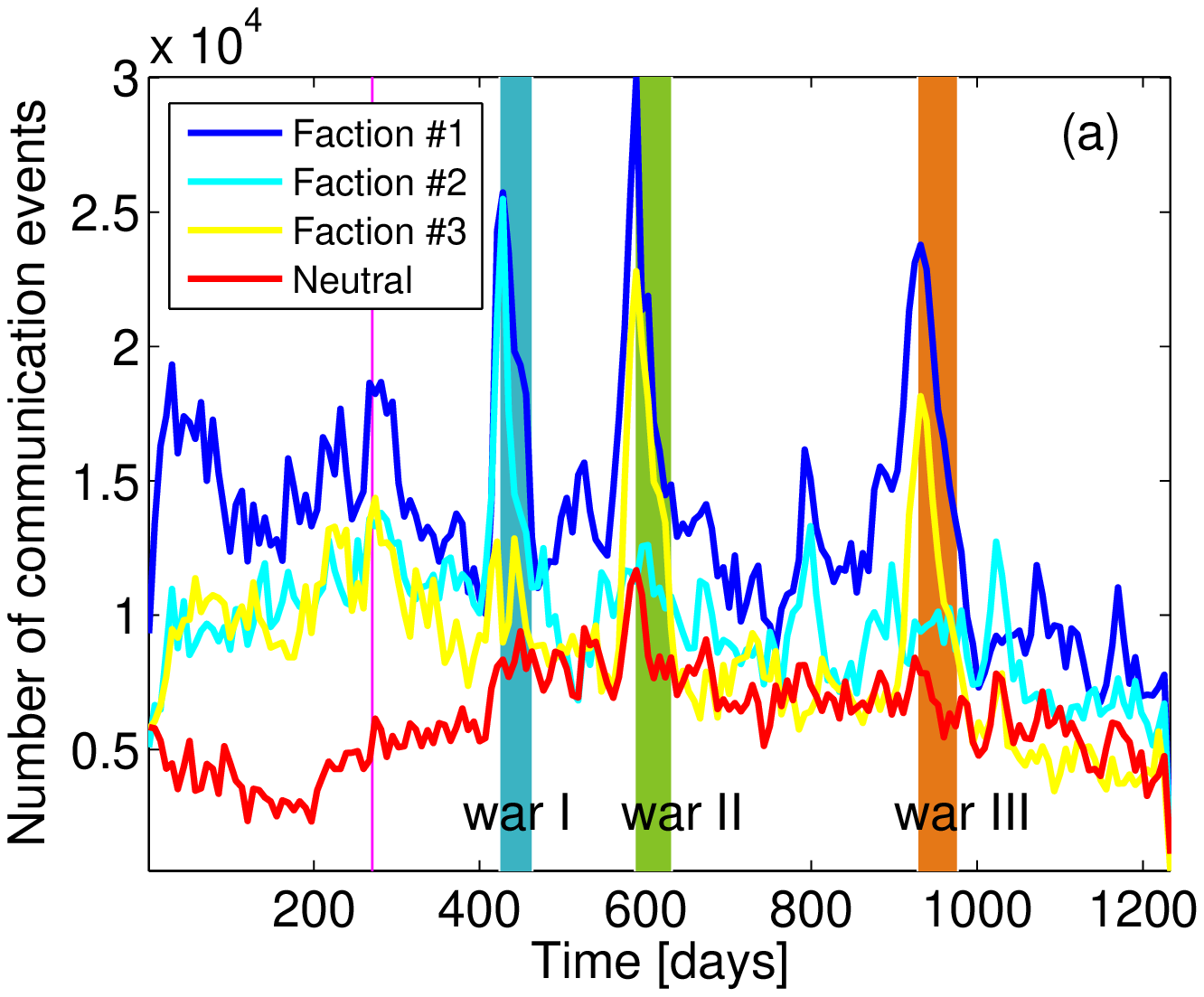}
\hspace{-3mm}
\includegraphics[width=0.55\textwidth]{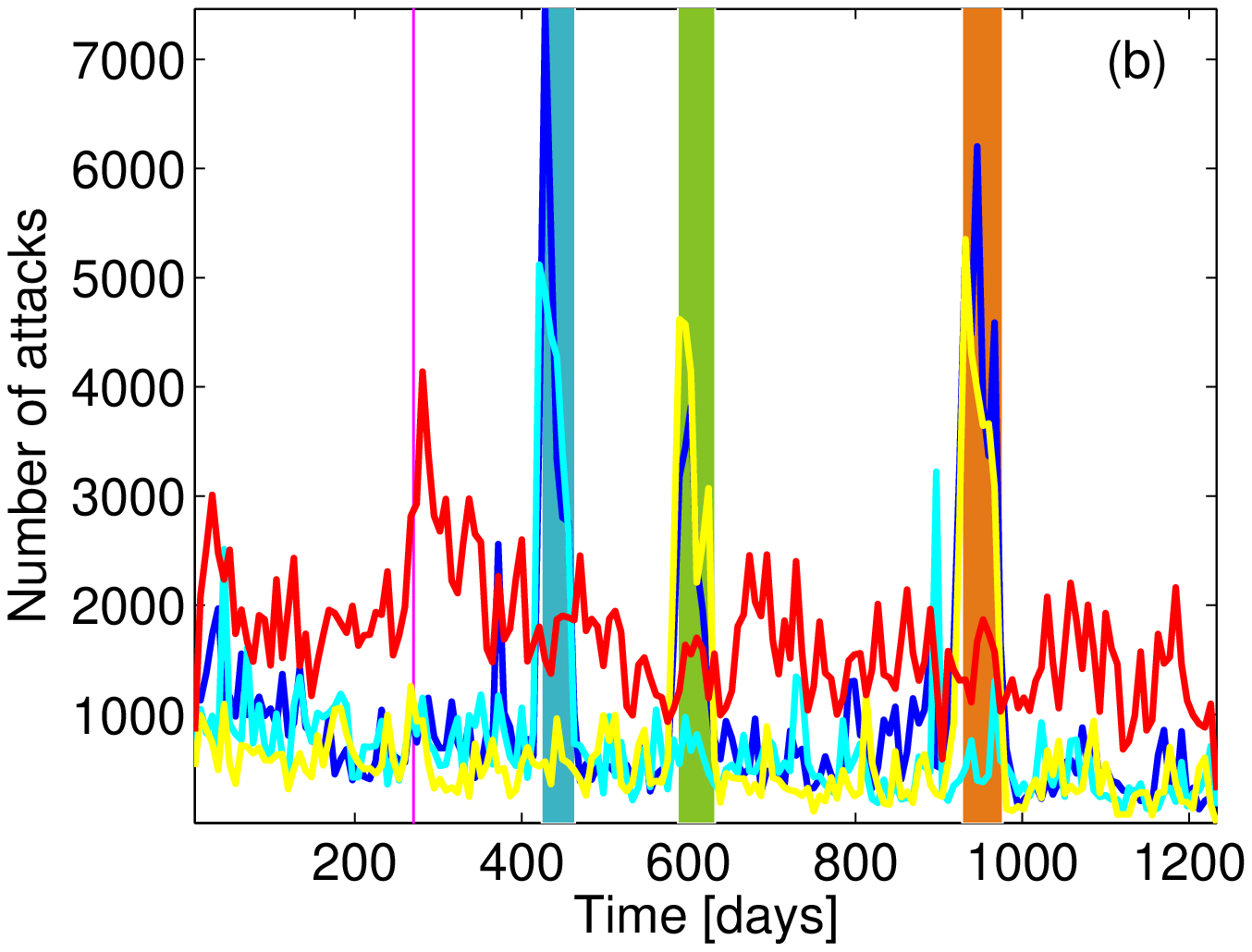}}
\caption{Number of weekly actions for different factions of players: (a) communication C; ( b) attacks A.
Activity of players involved in wars increased during the wars.
\label{fig10}}
\end{figure}

Table \ref{tab2} collects basic statistics of actions during war and peace periods. The change in activity that leads to the overall
increase of the number of actions of players engaged in war can be analyzed in more detail by studying the distributions for
specific actions. The most distinct peaks are observed for  C and A. In Fig. \ref{fig10} we show the weekly number of C
(a) and A (b) for the different factions.  The highest number of actions occurs in those factions that are involved in wars.
While the general level of activity during the wars increases, there are no distinct differences in the distributions of interevent times
for peace and war periods, respectively.

\begin{table}[]
\caption{Statistics of behavioural time series corresponding to war and peace periods (all actions).
$N$ is the number of players that performed actions in a given period;
$N_a$ is number of actions they performed; $N_\tau$ is the number of interevent time
($\tau$) values;
$\tau^{\rm max}$ is the maximal interevent time.  Also given,
percentage of positive and  negative actions, percentage of actions performed by male players. The last line
represents data for the overall peace period  (all war periods ignored).
 \label{tab2}}
{\footnotesize{ \centerline{
\begin{tabular}{l c r r r c c c }
\hline
&$N$& $N_a$&$N_\tau$&$\tau^{\rm max}$ [sec]&\parbox[t]{12mm}{\% pos actions}&\parbox[t]{12mm}{\% neg actions}&\% male\\
\hline
{All wars}$^{\dag}$ &4,883&1,350,144&1,345,119&9,856,569&84.9&15.1&87.5 \\
{War I}&
3,034&422,386&419,181&3,111,881&84.2&15.8&86.8\\
{War II}&
3,173&470,262&466,928&3,828,040&87.5&12.5&87.0\\
{War III}&
2,816&457,496&454,532&4,024,289&82.8&17.2&88.5\\
{Peace}$^{\ddag}$ &
7,776&7,023,065&7,015,281&52,305,174&90.4&9.6&86.4\\
\hline
\end{tabular}}}}
$^{\dag}$ Peace periods are ignored in the calculation of $t_{\mathrm{int}}$

$^{\ddag}$ War  periods are ignored in the calculation of $t_{\mathrm{int}}$
\end{table}

Taken that a war in a virtual world arises as a result of a complex process of social interactions between the players
it is tempting to seek for war precursors.
To this end we decided to check whether changes in player activity (of the different actions) may serve as a predictor for
an upcoming war. In {\em Pardus}Ê war officially starts whenever a so-called ``faction-relation'' ratio, defined by the game system for
each pair of factions, exceeds a certain threshold value. The faction-relation is evaluated daily in a specific way that takes into
account numerous behavioural factors of all players organised in factions, for more information see \cite{Pardus_web}.

To test if one can predict a war based on activity data only, we applied a cross-correlation analysis
and scanned for potential lead effects of activity patterns on the faction-relation time series.
We performed a systematic scan for lead-lag relationships, involving all types of actions (some are seen in Fig.~\ref{fig10}),
all combinations of factions, for several sizes of time-windows.
While we see clear differences between periods of war in peace in the extend of cross-correlations, we were unable to find conclusive
precursors for the onset of war.
This negative finding does not exclude the existence of other indicators that could have more predictive  power than the activity patterns alone.

\begin{figure}[t]
\centerline{\includegraphics[width=1.1\textwidth]{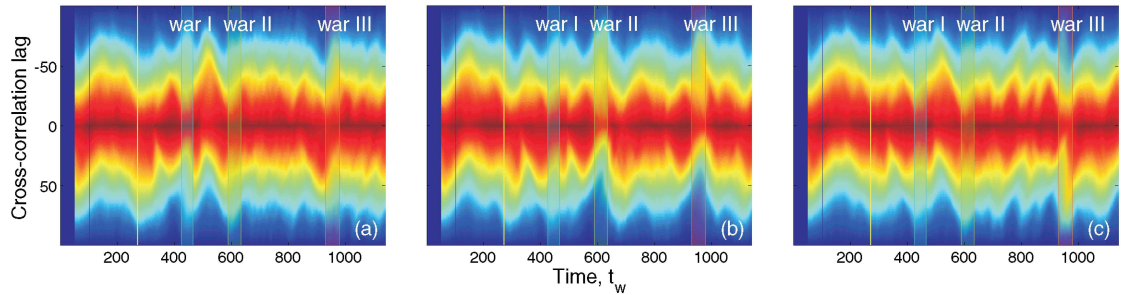}}
\caption{Cross-correlations between the number of
actions (all types pooled together) and the faction relation for
different factions using the sliding window of size 100 days. (a) factions 1 and 2. Note that they fight against each other in war I; (b)
factions 1 and 3. They  fight in wars II and III; (c) factions 2 and 3, who never fight against each other.
\label{fig13}}
\end{figure}

The cross-correlation values (colors) for all actions as a potential predictor for the faction-relation time series are presented in Fig. \ref{fig13}.
Lead and lag values are on positive and negative y-axis, respectively. War regions are marked by vertical bars.

\section{Conclusions}\label{V}

We used records about player activities in the MMOG {\em Pardus} to analyze several dynamical features of player actions.
It has been shown in previous work that the characters in {\em Pardus} display rich and realistic social behavior
\cite{Szell10a,Szell10b,Szell12,Thurner12,Szell13,Klimek13,Corominas13,Fuchs14a,Fuchs14b,Sinatra14}.
This includes the complex multiplex network structure of the {\em Pardus}
society,  complex multiplex network topologies of social interactions, non-ergodic behavioral action sequences, etc.
In the present study we extend these observations by a thorough analysis of the interevent time distributions $P_{\geq}(\tau)$
for the activities players can perform in the game.
In particular, we have shown that the interevent time distributions in the {\em Pardus} universe highly non-trivial
in nature. One feature of these distributions is an presence of periodic patterns on different
scales, which correspond on the one hand to trivial circadian cycles, on the other hand to more non-trivial periods of the working
day and to short scales of straightaway reactions. Similar to human activities in the real-world, non-trivial
distributions go hand in hand with bursty dynamics \cite{Barabasi05a,Oliveira05,Zhou08,Goh08}. The measured value of
burstiness of player actions, $B\simeq 0.6$ is compatible to the values reported for other types of human actions
\cite{Jo12,Yasseri12b} and implies long periods of inactivity between active periods.

Interevent time distributions  for different actions show that although these
distributions share several features,  they are specific for the various types of actions.
Different kinds of actions are characterized by different decay constants which indicate
the typical ``speed'' of performing: time intervals between messages are usually shorter
comparing with the other actions kinds, while the attack are  performed more fast than trade;
the longest  time is necessary for friendship and enmity marking.
We quantified decay times  by exponential, Eq.~(\ref{3}), and power law, Eq.~(\ref{2})  fits to $P_{\geq}(\tau)$,
depending on the time scale.
The interevent time distributions are well fitted by power law within the interval 6 min $< \tau <$ 8 hours,
while longer time intervals (6 min $< \tau <$ 8 hours) they are distributed exponentially.
For long interevent times (8 hours $< \tau < 24$) the differences between actions types become less pronounced.
The average values of interevent times $\overline{\tau}$ also indicates different speeds
in performing different types of actions. Even given the inactivity periods due to circadian cycles, the value of
$\overline{\tau}$ is smallest for communication and is much larger for other types.

Finally, we observed periods of increased activity in the {\em Pardus} due to history specific events, such as wars that were waged in the game.
Our search for potential predictive indicators that would signal the onset of war lead to the negative conclusion that changes in
player activity does not serve as reliable indicator of an upcoming massive conflict.

\section*{Acknowledgments}
This work was supported in part by the 7th FP, IRSES project No.~612707
Dynamics of and in Complex Systems (DIONICOS) and by Austrian
Science Fonds FWF~P23378.

\end{document}